\documentstyle[twoside]{article}

\catcode`\@=11
\long\def\@makefntext#1{ 
\protect\noindent \hbox to 3.2pt {\hskip-.9pt
$^{{\eightrm\@thefnmark}}$\hfil}#1\hfill} 

\def\thefootnote{\fnsymbol{footnote}}
 \def\@makefnmark{\hbox to 0pt{$^{\@thefnmark}$\hss}}  

\def\ps@myheadings{\let\@mkboth\@gobbletwo
\def\@oddhead{\hbox{} 
\rightmark\hfil\eightrm\thepage}
\def\@oddfoot{}\def\@evenhead{\eightrm\thepage\hfil 
\leftmark\hbox{}}\def\@evenfoot{}
\def\sectionmark##1{}\def\subsectionmark##1{}}

\oddsidemargin=\evensidemargin
\renewcommand{\thefootnote}{\fnsymbol{footnote}}

\newcounter{sectionc}\newcounter{subsectionc}\newcounter{subsubsectionc}
\renewcommand{\section}[1] {\vspace{12pt}\addtocounter{sectionc}{1}
\setcounter{subsectionc}{0}\setcounter{subsubsectionc}{0}\noindent
	{\tenbf\thesectionc. #1}\par\vspace{5pt}}
\renewcommand{\subsection}[1] {\vspace{12pt}\addtocounter{subsectionc}{1}
	\setcounter{subsubsectionc}{0}\noindent
	{\bf\thesectionc.\thesubsectionc. {\kern1pt \bfit #1}}\par\vspace{5pt}}
\renewcommand{\subsubsection}[1] {\vspace{12pt}\addtocounter{subsubsectionc}{1}
	\noindent{\tenrm\thesectionc.\thesubsectionc.\thesubsubsectionc.
	{\kern1pt \tenit #1}}\par\vspace{5pt}}
\newcommand{\nonumsection}[1] {\vspace{12pt}\noindent{\tenbf #1}
	\par\vspace{5pt}}

\newcounter{appendixc}
\newcounter{subappendixc}[appendixc]
\newcounter{subsubappendixc}[subappendixc]
\renewcommand{\thesubappendixc}{\Alph{appendixc}.\arabic{subappendixc}}
\renewcommand{\thesubsubappendixc}
	{\Alph{appendixc}.\arabic{subappendixc}.\arabic{subsubappendixc}}

\renewcommand{\appendix}[1] {\vspace{12pt}
        \refstepcounter{appendixc}
        \setcounter{figure}{0}
        \setcounter{table}{0}
        \setcounter{lemma}{0}
        \setcounter{theorem}{0}
        \setcounter{corollary}{0}
        \setcounter{definition}{0}
        \setcounter{equation}{0}
        \renewcommand{\thefigure}{\Alph{appendixc}.\arabic{figure}}
        \renewcommand{\thetable}{\Alph{appendixc}.\arabic{table}}
        \renewcommand{\theappendixc}{\Alph{appendixc}}
        \renewcommand{\thelemma}{\Alph{appendixc}.\arabic{lemma}}
        \renewcommand{\thetheorem}{\Alph{appendixc}.\arabic{theorem}}
        \renewcommand{\thedefinition}{\Alph{appendixc}.\arabic{definition}}
        \renewcommand{\thecorollary}{\Alph{appendixc}.\arabic{corollary}}
        \renewcommand{\theequation}{\Alph{appendixc}.\arabic{equation}}
        \noindent{\tenbf Appendix \theappendixc #1}\par\vspace{5pt}}
\newcommand{\subappendix}[1] {\vspace{12pt}
        \refstepcounter{subappendixc}
        \noindent{\bf Appendix \thesubappendixc. {\kern1pt \bfit #1}}
	\par\vspace{5pt}}
\newcommand{\subsubappendix}[1] {\vspace{12pt}
        \refstepcounter{subsubappendixc}
        \noindent{\rm Appendix \thesubsubappendixc. {\kern1pt \tenit #1}}
	\par\vspace{5pt}}

\newcommand{\textlineskip}{\baselineskip=13pt}
\newcommand{\smalllineskip}{\baselineskip=10pt}

\def\eightcirc{
\begin{picture}(0,0)
\put(4.4,1.8){\circle{6.5}}
\end{picture}}
\def\eightcopyright{\eightcirc\kern2.7pt\hbox{\eightrm c}}

\newcommand{\copyrightheading}[1]
	{\vspace*{-2.5cm}\smalllineskip{\flushleft
	{\eightrm Modern Physics Letters B, #1}\\
	{\eightrm $\eightcopyright$\, World Scientific Publishing
	 Company}\\
	 }}

\newcommand{\pub}[1]{{\begin{center}\eightrm\smalllineskip
	Submitted #1\\
	\end{center}
	}}

\def\abstracts#1#2#3{{
	\centering{\begin{minipage}{4.5in}\baselineskip=10pt\eightrm
	\centerline{ABSTRACT}
	\parindent=0pt #1\par
	\parindent=15pt #2\par
	\parindent=15pt #3
	\end{minipage} }\par}}

\newcommand{\bibit}{\nineit}
\newcommand{\bibbf}{\ninebf}
\renewenvironment{thebibliography}[1]			
	{\ninerm
	 \baselineskip=11pt				
	 \begin{list}{\arabic{enumi}.}
	{\usecounter{enumi}\setlength{\parsep}{0pt}
	 \setlength{\leftmargin 17pt}{\rightmargin 0pt}	
	 \setlength{\itemsep}{0pt} \settowidth		
	{\labelwidth}{#1.}\sloppy}}{\end{list}}

\newcounter{itemlistc}
\newcounter{romanlistc}
\newcounter{alphlistc}
\newcounter{arabiclistc}

\newenvironment{romanlist}
	{\setcounter{romanlistc}{0}
	 \begin{list}{$($\roman{romanlistc}$)$}
	{\usecounter{romanlistc}
	 \setlength{\parsep}{0pt}
	 \setlength{\itemsep}{0pt}}}{\end{list}}

\newcommand{\fcaption}[1]{
        \refstepcounter{figure}
        \setbox\@tempboxa = \hbox{\eightrm Fig.~\thefigure. #1}
        \ifdim \wd\@tempboxa > 5in
           {\begin{center}
        \parbox{5in}{\eightrm \smalllineskip Fig.~\thefigure. #1 }
            \end{center}}
        \else
             {\begin{center}
             {\eightrm Fig.~\thefigure. #1}
              \end{center}}
        \fi}

\newcommand{\tcaption}[1]{
        \refstepcounter{table}
        \setbox\@tempboxa = \hbox{\eightrm Table~\thetable. #1}
        \ifdim \wd\@tempboxa > 5in
           {\begin{center}
        \parbox{5in}{\eightrm\smalllineskip Table~\thetable. #1 }
            \end{center}}
        \else
             {\begin{center}
             {\eightrm Table~\thetable. #1}
              \end{center}}
        \fi}

\def\@citex[#1]#2{\if@filesw\immediate\write\@auxout	
	{\string\citation{#2}}\fi			
\def\@citea{}\@cite{\@for\@citeb:=#2\do			
	{\@citea\def\@citea{,}\@ifundefined		
	{b@\@citeb}{{\bf ?}\@warning
	{Citation `\@citeb' on page \thepage \space undefined}}
	{\csname b@\@citeb\endcsname}}}{#1}}

\newif\if@cghi
\def\cite{\@cghitrue\@ifnextchar [{\@tempswatrue
	\@citex}{\@tempswafalse\@citex[]}}
\def\citelow{\@cghifalse\@ifnextchar [{\@tempswatrue
	\@citex}{\@tempswafalse\@citex[]}}
\def\@cite#1#2{{$\null^{#1}$\if@tempswa\typeout
	{IJCGA warning: optional citation argument
	ignored: `#2'} \fi}}

\def\pmb#1{\setbox0=\hbox{#1}
	\kern-.025em\copy0\kern-\wd0
	\kern.05em\copy0\kern-\wd0
	\kern-.025em\raise.0433em\box0}

\def\fnt#1#2{\footnotetext{\kern-.3em
	{$^{\mbox{\scriptsize #1}}$}{#2}}}

\def\fpage#1{\begingroup
\voffset=.3in
\thispagestyle{empty}\begin{table}[b]\centerline{\footnotesize #1}
	\end{table}\endgroup}

\def\runninghead#1#2{\pagestyle{myheadings}
\markboth{{\eightit{\quad #1}}\hfill}{\hfill{\eightit{#2\quad}}}}

\font\tenbf=cmbx10
\font\tenit=cmti10
\font\tenit=cmti10
\font\bfit=cmbxti10 at 10pt
 1
 1
 1
\font\ninebf=cmbx9
\font\ninerm=cmr9
\font\nineit=cmti9

\font\eightrm=cmr8
\font\eightit=cmti8

\def\qed{\hbox{${\vcenter{\vbox{                          
   \hrule height 0.4pt\hbox{\vrule width 0.4pt height 6pt
   \kern5pt\vrule width 0.4pt}\hrule height 0.4pt}}}$}}

\runninghead{C. W. J. Beenakker} {Exactly Solvable Scaling
Theory of Conduction in Disordered Wires}

\begin{document}
\normalsize\textlineskip
{\thispagestyle{empty}
\setcounter{page}{1}
\renewcommand{\thefootnote}{\fnsymbol{footnote}} 
\copyrightheading{to be published as ``Brief Review''
--- cond-mat/9403033}
\vspace*{0.88truein}
\fpage{1}

\centerline{\bf EXACTLY SOLVABLE SCALING THEORY}
\vspace*{0.035truein}
\centerline{\bf OF CONDUCTION IN DISORDERED WIRES}
\vspace{0.37truein}
\centerline{\footnotesize C. W. J. Beenakker}
\vspace*{0.015truein}
\centerline{\footnotesize\it Instituut-Lorentz, University of Leiden}
\baselineskip=10pt
\centerline{\footnotesize\it P.O. Box 9506, 2300 RA Leiden, The Netherlands}
\vspace{0.225truein}
\pub{March 1994}
\vspace*{0.21truein}

\abstracts{\noindent
Recent developments are reviewed in the scaling theory of
phase-coherent conduction through a disordered wire. The
Dorokhov-Mello-Pereyra-Kumar equation for the distribution of
transmission eigenvalues has been solved exactly, in the absence of
time-reversal symmetry. Comparison with the previous prediction of
random-matrix theory shows that this prediction was highly accurate ---
but not exact:  The repulsion of the smallest eigenvalues was
overestimated by a factor of two. This factor of two resolves several
disturbing discrepancies between random-matrix theory and
microscopic calculations, notably in the magnitude of the universal
conductance fluctuations in the metallic regime, and in the width of
the log-normal conductance distribution in the insulating regime.
}{}{}

\vspace*{-3pt}\textlineskip
\section{Introduction}
\noindent
In 1980, Anderson, Thouless, Abrahams, and Fisher\cite{And80} proposed
a {\em ``new method for a scaling theory of localization''}, based on
Landauer's interpretation of electrical conduction as quantum
mechanical transmission.\cite{Lan57} They considered a one-dimensional
(1D) chain with weak scattering (mean free path $l$ much greater than
the Fermi wave length $\lambda_{\rm F}$), and computed how the
transmission probability $T$ scales with the chain length $L$. For
$L>l$ an exponential decay was obtained, demonstrating localization. In
the following decade the scaling theory of 1D localization was
developed in great detail,$^{3-7}$ and the complete distribution
$P(T,L)$ of the transmission probability was found (and hence of the
conductance $G=T\times 2e^{2}/h$). One can thus regard the problem of
1D localization as solved, at least for the case of weak scattering.

A real metal wire is not one-dimensional. Typically, the width $W$ is
much greater than $\lambda_{\rm F}$, so that the number $N$ of
transverse modes at the Fermi level is much greater than one. Instead
of a single transmission probability $T$, one now has $N$ transmission
eigenvalues $T_{1},T_{2},\ldots T_{N}$. (The numbers $T_{n}\in[0,1]$
are the eigenvalues of the matrix product $tt^{\dagger}$, where $t$ is
the $N\times N$ transmission matrix of the wire.) To obtain the
distribution of the conductance
\begin{equation}
G=\frac{2e^{2}}{h}\sum_{n=1}^{N}T_{n},\label{Landauer}
\end{equation}

\eject}
\textheight=7.8truein
\setcounter{footnote}{0}
\renewcommand{\thefootnote}{\alph{footnote}}

\noindent
one now needs the joint probability distribution $P(T_{1},T_{2},\ldots
T_{N},L)$. This distribution differs essentially from the distribution
in the 1D chain, because of strong correlations between the
transmission eigenvalues. These correlations originate from a
``repulsion'' of nearby eigenvalues. As a consequence of this
eigenvalue repulsion, the localization length is increased by a factor
of $N$ in comparison to the 1D case. One can therefore distinguish a
metallic and an insulating regime. On length scales $l<L<Nl$ the
conductance decreases linearly rather than exponentially with $L$. This
is the (diffusive) metallic regime, where mesoscopic effects as weak
localization and universal conductance fluctuations (UCF) occur. The
insulating regime of exponentially small conductance is entered for
wire lengths $L>Nl$.

A scaling theory of localization in multi-mode wires was pioneered by
Dorokhov,\cite{Dor82} and independently by Mello, Pereyra, and
Kumar.\cite{Mel88} The DMPK scaling equation,
\begin{eqnarray}
l\frac{\partial P}{\partial L}&=&
\frac{2}{\beta N+2-\beta}\sum_{i=1}^{N}
\frac{\partial}{\partial\lambda_{i}}\lambda_{i}(1+\lambda_{i})
J\frac{\partial}{\partial\lambda_{i}}J^{-1}P,\label{DMPK}\\
J&=&\prod_{i<j}|\lambda_{j}-\lambda_{i}|^{\beta},
\label{jacobian}
\end{eqnarray}
describes the evolution of the distribution function
$P(\lambda_{1},\lambda_{2},\ldots \lambda_{N},L)$ in an ensemble of
disordered wires of increasing length. The variables
$\lambda_{n}\in[0,\infty)$ are simply related to the transmission
eigenvalues by $\lambda_{n}\equiv(1-T_{n})/T_{n}$. The ensemble is
characterized by a mean free path $l$ and by a symmetry index $\beta$,
which takes on the values 1, 2, and 4 depending on the presence or
absence of time-reversal and spin-rotation symmetry ($\beta=2$ in the
presence of a magnetic field; otherwise, $\beta=1$ or 4 in the absence
or presence of spin-orbit scattering).

The DMPK equation has the form of a diffusion equation in eigenvalue
space. The function $J$ which couples the degrees of
freedom is the Jacobian from the space of scattering matrices to the
space of transmission eigenvalues. The similarity to diffusion in real
space has been given further substance by the
demonstration\cite{Mel88c} that Eq.\ (\ref{DMPK}) holds on length
scales $\gg l$ regardless of the microscopic scattering properties of
the conductor ({\em one-parameter scaling}).  Eq.\ (\ref{DMPK}) was
derived by Dorokhov,\cite{Dor82} (for $\beta=2$) and by Mello, Pereyra,
and Kumar,\cite{Mel88} (for $\beta=1$, with generalizations to
$\beta=2,4$ in Refs.\ 11,12) by computing the incremental change of the
transmission eigenvalues upon attachment of a thin slice to the wire.
It is assumed that the conductor is weakly disordered
($l\gg\lambda_{\rm F}$), so that the scattering in the thin slice can
be treated by perturbation theory. A key simplification is the isotropy
assumption that the flux incident in one scattering channel is, on
average, equally distributed among all outgoing channels. This
assumption restricts the applicability of the DMPK equation to a wire
geometry ($L\gg W$), since it ignores the finite time scale for
transverse diffusion.

The DMPK equation has been studied extensively for more than ten years.
The strong coupling of the scattering channels by the Jacobian
prevented an exact solution by standard methods. The problem simplifies
drastically deep in the localized regime ($L\gg Nl$), when the
scattering channels become effectively decoupled. Pichard\cite{Pic91}
has computed from Eq.\ (\ref{DMPK}) the log-normal distribution of the
conductance in this regime, and has found an excellent agreement with
numerical simulations. In the metallic regime ($L\ll Nl$), Mello and
Stone\cite{Mel91,Mel88b} were able to compute the first two moments of
the conductance, in precise agreement with the diagrammatic
perturbation theory of weak localization and UCF.  More general
calculations of the weak localization effect\cite{Bee94} and of
universal fluctuations\cite{Cha93} [for arbitrary transport properties
of the form $A=\sum_{n}f(T_{n})$] have been developed, based on
linearization of Eq.\ (\ref{DMPK}) in the fluctuations of the
$\lambda$'s around their mean positions (valid in the large-$N$
metallic regime, when the fluctuations are small). None of these
approaches was capable of finding the full distribution function. The
purpose of this paper is to review some recent work by B. Rejaei and
the author,\cite{Rej93} in which the DMPK scaling equation has been
solved {\em exactly\/} for the case $\beta=2$.

\section{Random-matrix theory and the 1/8 --- 2/15 puzzle}
\noindent
There existed a special and urgent reason for wanting the
full distribution function of the transmission eigenvalues. We are
referring to a disturbing discrepancy\cite{Bee93} between the
random-matrix theory of UCF and the established diagrammatic
perturbation theory. In order to appreciate the significance of the
recent developments, it seems worthwhile to discuss this issue in some
``historical'' perspective.

In the sixties, Wigner, Dyson, Mehta, and others developed
random-matrix theory (RMT) into a powerful tool to study the statistics
of energy levels measured in nuclear reactions.\cite{Por65} It was
shown that the fluctuations in the energy level density are governed by
level repulsion. Mathematically, level repulsion originates from the
Jacobian $J=\prod_{i<j}|E_{j}-E_{i}|^{\beta}$ of the transformation
from matrix space to eigenvalue space, which depends on the symmetry of
the Hamiltonian ensemble --- but is independent of the mean level
density.\cite{Meh67} This universality is at the origin of the
remarkable success of RMT in nuclear physics.\cite{Bro81}  The
universality of the level fluctuations is expressed by the celebrated
Dyson-Mehta formula\cite{Dys63} for the variance of a linear statistic
$A=\sum_{n}a(E_{n})$ on the energy levels $E_{n}$. (The quantity $A$ is
called a linear statistic because products of different $E_{n}$'s do
not appear, but the function $a(E)$ may well depend non-linearly on
$E$.) The Dyson-Mehta formula reads
\begin{equation}
{\rm Var}\,A=\frac{1}{\beta}\,\frac{1}
{\pi^{2}}\int_{0}^{\infty}\!\!dk\,|a(k)|^{2}k,\label{DysonMehta}
\end{equation}
where $a(k)=\int_{-\infty}^{\infty}\!dE\,{\rm e}^{{\rm i}kE}a(E)$ is
the Fourier transform of $a(E)$.  Eq.\ (\ref{DysonMehta}) shows that:
1.~The variance is independent of microscopic parameters; 2.~The
variance has a universal $1/\beta$-dependence on the symmetry index.

In a seminal 1986-paper,\cite{Imr86} Imry proposed to apply RMT to the
phenomenon of universal conductance fluctuations, which was discovered
using diagrammatic perturbation theory by Al'tshuler\cite{Alt85} and
Lee and Stone.\cite{Lee85} UCF is the occurrence of sample-to-sample
fluctuations in the conductance which are of order $e^{2}/h$ at zero
temperature, {\em independent\/} of the size of the sample or the
degree of disorder --- as long as the conductor remains in the diffusive
metallic regime.  The relationship between the statistics of energy
levels measured in nuclear reactions on the one hand, and the
statistics of conductance fluctuations measured in transport
experiments on the other hand, was used by Muttalib, Pichard, and
Stone\cite{Mut87} to develop a random-matrix theory of quantum
transport. (For a review, see Ref.\ 27.) The RMT of quantum
transport differs from the RMT of level statistics in two essential
ways.

\begin{romanlist}
\item
The first is that the transmission eigenvalues $T_{n}$ are {\em not\/}
the eigenvalues of the scattering matrix. Instead they are the
eigenvalues of $tt^{\dagger}$, where the transmission matrix $t$ is an
$N\times N$ submatrix of the $2N\times 2N$ scattering matrix of the
conductor. It turns out that the repulsion of the variables
$\lambda_{n}\equiv(1-T_{n})/T_{n}$ takes the same form as the repulsion
of the energy levels $E_{n}$. More precisely, the Jacobian
(\ref{jacobian}) in terms of the $\lambda$'s has the same form as for
level statistics.  Random-matrix theory is based on the fundamental
assumption that all correlations between the eigenvalues are due to the
Jacobian. If all correlations are due to the Jacobian, then the
probability distribution $P(\lambda_{1},\lambda_{2},\ldots\lambda_{N})$
of the $\lambda$'s should have the form $P\propto
J\prod_{i}p(\lambda_{i})$, or equivalently,
\begin{eqnarray}
P(\{\lambda_{n}\})&=&C\exp\Bigl[-\beta\Bigl(\sum_{i<j}
u(\lambda_{i},\lambda_{j})+\sum_{i}V(\lambda_{i})\Bigr)\Bigr],
\label{Pglobala}\\
u(\lambda_{i},\lambda_{j})&=&-\ln|\lambda_{j}-\lambda_{i}|,
\label{Pglobalb}
\end{eqnarray}
with $V=-\beta^{-1}\ln p$ and $C$ a normalization constant.
Eq.\ (\ref{Pglobala}) has the form of a Gibbs distribution at
temperature $\beta^{-1}$ for a fictitious system of classical particles
on a line in an external potential $V$, with a logarithmically
repulsive interaction $u$. All microscopic parameters are contained in
the single function $V(\lambda)$. The logarithmic repulsion is
independent of microscopic parameters, because of its geometric
origin.

\item
The second difference is that the correlation function of the
$\lambda$'s is not translationally invariant, due to the positivity
constraint on $\lambda$. This constraint $\lambda\geq 0$ follows
directly from unitarity of the scattering matrix. In contrast, the
correlation function in the RMT of energy levels is translationally
invariant over the energy range of interest. Because of this
complication, it could not be shown that the universality of the
fluctuations is generic for arbitrary linear statistics on the
transmission eigenvalues. In particular, no formula with the generality
of the Dyson-Mehta formula (\ref{DysonMehta}) could be derived. The
lack of such a general theory was being felt especially since
mesoscopic fluctuations in transport properties other than the
conductance (both in conductors and superconductors) became of
interest. Examples are the critical-current fluctuations in Josephson
junctions,\cite{Bee91} conductance fluctuations at
normal-superconductor interfaces,\cite{Tak91} and fluctuations in the
shot-noise power of metals.\cite{Jon92} This obstacle towards the
establishment of universality in the RMT of quantum transport was
finally overcome in 1993,\cite{Bee93} when a technique was developed to
compute correlation functions by a method of functional derivatives,
which does not require translational invariance. The analogue could be
obtained of the Dyson-Mehta formula for the variance of a linear
statistic $A=\sum_{n}f(T_{n})$ on the transmission eigenvalues:
\begin{equation}
{\rm Var}\,A=\frac{1}{\beta}\,\frac{1}{\pi^{2}}\int_{0}^{\infty}
\!\!dk\,|F(k)|^{2}k\tanh(\pi k).\label{CB}
\end{equation}
The function $F(k)$ is defined in terms of the function $f(T)$ by the
transform
\begin{equation}
F(k)=\int_{-\infty}^{\infty}\!dx\,{\rm e}^{{\rm i}kx}f
\left(\frac{1}{1+{\rm e}^{x}}\right).\label{Fkdef}
\end{equation}
The formula (\ref{CB}) demonstrates that the universality which was the
hallmark of UCF is generic for a whole class of transport properties,
viz.\ those which are linear statistics on the transmission
eigenvalues.
\end{romanlist}

The probability distribution (\ref{Pglobala}) was justified by a
maximum-entropy principle for quasi-1D conductors.\cite{Mut87,Sto91}
Quasi-1D means $L\gg W\gg\lambda_{\rm F}$. In this limit one can assume
that the distribution of scattering matrices is only a function of the
transmission eigenvalues (isotropy assumption). The distribution
(\ref{Pglobala}) then maximizes the information entropy subject to the
constraint of a given density of eigenvalues. The function $V(\lambda)$
is determined by this constraint and is not specified by RMT.

It was initially believed that Eq.\ (\ref{Pglobala}) would provide an
exact description in the quasi-1D limit, if only $V(\lambda)$ were
suitably chosen.\cite{Sto91} However, the generalized Dyson-Mehta
formula (\ref{CB}) demonstrates that RMT is not exact, not even in the
quasi-1D limit. If one computes from Eq.\ (\ref{CB}) the variance of
the conductance (\ref{Landauer}) [by substituting $f(T)=G_{0}T$, with
$G_{0}=2e^{2}/h$], one finds
\begin{equation}
{\rm Var\,}G/G_{0}=\frac{1}{8}\beta^{-1},\label{UCFglobal}
\end{equation}
independent of the form of $V(\lambda)$. The diagrammatic perturbation
theory\cite{Alt85,Lee85} of UCF gives instead
\begin{equation}
{\rm Var\,}G/G_{0}=\frac{2}{15}\beta^{-1}\label{UCF}
\end{equation}
for a quasi-1D conductor. The difference between the coefficients
$\frac{1}{8}$ and $\frac{2}{15}$ is tiny, but it has the fundamental
implication that the interaction between the $\lambda$'s is not
precisely logarithmic, or in other words, that there exist correlations
between the transmission eigenvalues over and above those induced by
the Jacobian.

The $\frac{1}{8}$~---~$\frac{2}{15}$ discrepancy raised the question
what the true eigenvalue interaction would be in quasi-1D conductors.
Is there perhaps a cutoff for large separation of the $\lambda$'s? Or
is the true interaction a many-body interaction, which can not be
reduced to the sum of pairwise interactions? This transport problem has
a counterpart in a closed system. The RMT of the statistics of the
eigenvalues of a random Hamiltonian yields a probability distribution
of the form (\ref{Pglobala}), with a logarithmic repulsion between the
energy levels.\cite{Meh67} It was shown by Efetov\cite{Efe83} and by
Al'tshuler and Shklovski\u{\i}\cite{Alt86} that the logarithmic
repulsion in a small disordered particle (length $L$, diffusion
constant $D$) holds for energy separations small compared to the
Thouless energy $E_{\rm c}\equiv\hbar D/L^{2}$. For larger separations
the interaction potential decays algebraically.\cite{Jal93} As we shall
discuss, the way in which the RMT of quantum transport breaks down is
quite different.

\section{Non-logarithmic eigenvalue repulsion}
\noindent
The method of solution of the DMPK equation is a mapping onto a model
of non-interacting fermions, inspired by Sutherland's mapping of a
different diffusion equation.\cite{Sut72} The case $\beta=2$ is
special, because for other values of $\beta$ the mapping introduces
interactions between the fermions. The free-fermion problem has the
character of a one-dimensional scattering problem in imaginary time,
which can be solved exactly without great difficulties. The reader who
is interested in ``how it is done'' is referred to
Ref.\ 17. In this {\em Brief Review\/} we limit ourselves
to presenting the solution and discussing its implications.

The DMPK equation (\ref{DMPK}) [with $\beta=2$] can be solved for
arbitrary initial conditions. We consider the  ballistic initial
condition $\lim_{L\rightarrow 0}P=\prod_{i}\delta(\lambda_{i})$,
appropriate for the case of ideal contacts. The solution is given by
the square root of the Jacobian (\ref{jacobian}) times the determinant
of an $N$-dimensional matrix $M$. The determinant is the Slater
determinant of the free-fermion problem. The square root of the
Jacobian comes from the mapping of the DMPK equation onto the
Schr\"{o}dinger equation. The solution is
\begin{eqnarray}
&&P(\{\lambda_{n}\},L)=C(L)\,J^{1/2}\,|{\rm Det}\,M|,\label{Psolution}\\
&&M_{nm}=\int_{0}^{\infty}\!\!dk\,
\exp(-{\textstyle\frac{1}{4}}k^{2}L/Nl)
\tanh({\textstyle\frac{1}{2}}\pi k)k^{2m-1}\,
{\rm P}_{\frac{1}{2}({\rm i}k-1)}(1+2\lambda_{n}),\label{Msolution}
\end{eqnarray}
where $C(L)$ is a $\lambda$-independent normalization factor. Using
an integral representation for the Legendre functions ${\rm P}_{\nu}$,
the matrix elements (\ref{Msolution}) can be rewritten in terms of
Hermite polynomials
${\rm H}_{2m-1}$,
\begin{equation}
M_{nm}=c\int_{{\rm arccosh}(1+2\lambda_{n})}^{\infty}
\!\!\!\!\!\!du\,
\exp(-{\textstyle\frac{1}{4}}u^{2}Nl/L)
(\cosh u-1-2\lambda_{n})^{-1/2}
{\rm H}_{2m-1}({\textstyle\frac{1}{2}}u\sqrt{Nl/L}),\label{Msolution2}
\end{equation}
where $c$ is another constant which can be absorbed in $C(L)$.

For $N=1$, the Jacobian $J\equiv 1$ and ${\rm Det}\,M=M_{11}$, so that
Eq.\ (\ref{Psolution}) reduces to
\begin{equation}
P(\lambda,L)=C(L)\int_{{\rm arccosh}(1+2\lambda)}^{\infty}
\!\!\!\!\!du\,
\exp(-{\textstyle\frac{1}{4}}u^{2}l/L)
(\cosh u-1-2\lambda)^{-1/2}\,u.\label{PN1}
\end{equation}
Normalization gives
$C(L)=(2\pi)^{-1/2}(l/L)^{3/2}\exp(-\frac{1}{4}L/l)$.  This is
Abrikosov's solution\cite{Abr81} of the scaling equation for a 1D
chain.\footnote{
This solution (\protect\ref{PN1}) of the 1D scaling equation was
actually obtained as early as 1959 by Gertsenshtein and
Vasil'ev,\cite{Ger59} in a paper entitled {\em ``Waveguides with random
inhomogeneities and Brownian motion in the Lobachevsky plane.''} This
remarkable paper on the exponential decay of radio-waves due to weak
disorder contains many of the results which were rederived in the
eighties for the problem of 1D localization of electrons.$^{3-7}$ The
paper was noticed in the optical literature,\cite{Pap71} but apparently
not among solid-state physicists.
} This solution is $\beta$-independent ($\beta$ drops out of
Eq.\ (\ref{DMPK}) for $N=1$). Equation (\ref{Psolution}) generalizes
the 1D-chain solution to arbitrary $N$, for the case $\beta=2$.

The Slater determinant can be evaluated in closed form in the metallic
regime $L\ll Nl$ and in the insulating regime $L\gg Nl$. In both
regimes the probability distribution takes the form (\ref{Pglobala}) of
a Gibbs distribution with a parameter-independent two-body interaction
$u(\lambda_{i},\lambda_{j})$, as predicted by RMT. However, the
interaction differs from the logarithmic repulsion (\ref{Pglobalb}) of
RMT. Instead, it is given by\cite{Rej93}
\begin{equation}
u(\lambda_{i},\lambda_{j})=-{\textstyle\frac{1}{2}}
\ln|\lambda_{j}-\lambda_{i}|
-{\textstyle\frac{1}{2}}\ln|{\rm arcsinh}^{2}\lambda_{j}^{1/2}-{\rm
arcsinh}^{2}\lambda_{i}^{1/2}|.\label{final3u}
\end{equation}
The eigenvalue interaction (\ref{final3u}) is different for weakly and
for strongly transmitting scattering channels:
$u\rightarrow-\ln|\lambda_{j}-\lambda_{i}|$ for
$\lambda_{i},\lambda_{j}\ll 1$, but
$u\rightarrow-\frac{1}{2}\ln|\lambda_{j}-\lambda_{i}|$ for
$\lambda_{i},\lambda_{j}\gg 1$. For weakly transmitting channels it is
{\em twice as small\/} as predicted by considerations based solely on
the Jacobian, which turn out to apply only to the strongly transmitting
channels.

In the metallic regime $L\ll Nl$, the method of functional derivatives
of Ref.\ 18 can still be used to compute the variance of a linear
statistic, since this method works for any two-body interaction.
Instead of Eq.\ (\ref{CB}), one now obtains for the variance the
formula
\begin{eqnarray}
&&{\rm Var\,}A=\frac{1}{\beta}\,\frac{1}{\pi^{2}}\int_{0}^{\infty}\!\!dk\,
\frac{k|F(k)|^{2}}{1+{\rm cotanh}
({\textstyle\frac{1}{2}}\pi k)},\label{VarAresult}\\
&&F(k)=\int_{-\infty}^{\infty}\!\!dx\,{\rm e}^{{\rm i}kx}
f\left(\frac{1}{\cosh^{2}x}\right).\label{akdef}
\end{eqnarray}
This result was obtained for $\beta=2$ from the exact solution given
above,\cite{Rej93} and independently for all $\beta\in\{1,2,4\}$ by the
perturbative method of Chalker and Mac\^{e}do.\cite{Cha93} Substitution
of $f(T)=T$ now yields $\frac{2}{15}$ instead of $\frac{1}{8}$ for the
coefficient of the UCF, thus resolving the discrepancy between
Eqs.\ (\ref{UCFglobal}) and (\ref{UCF}). The conclusion is that the
discrepancy with RMT originated from a reduced repulsion of weakly
transmitting channels.

In the insulating regime $L\gg Nl$, all $\lambda$'s are exponentially
large, and the interaction (\ref{final3u}) may be effectively
simplified by
$u(\lambda_{i},\lambda_{j})=-\frac{1}{2}\ln|\lambda_{j}-\lambda_{i}|$.
This is a factor of two smaller than the interaction (\ref{Pglobalb})
predicted by RMT. This explains the factor-of-two discrepancy between
the results of RMT and of numerical simulations for the width of the
log-normal distribution of the conductance:\cite{Pic91} RMT predicts
${\rm Var}\,\ln G/G_{0}=-\langle\ln G/G_{0}\rangle$, which is twice as
small as the result
\begin{equation}
{\rm Var}\,\ln G/G_{0}=-2\langle\ln G/G_{0}\rangle\label{VarlnG}
\end{equation}
which follows from the exact solution of the DMPK equation for
$\beta=2$. As shown by Pichard,\cite{Pic91} the relationship
(\ref{VarlnG}) between mean and variance of $\ln G/G_{0}$ remains valid
for other values of $\beta$, since both the mean and the variance have
a $1/\beta$ dependence on the symmetry index.

\newpage
\section{Outlook}
\noindent
We conclude by mentioning some directions for future research. So far
only the case $\beta=2$ of broken time-reversal symmetry has been
solved exactly.\cite{Rej93} In that case the DMPK equation (\ref{DMPK})
can be mapped onto a free-fermion problem. For $\beta=1,4$ the
Sutherland-type mapping is onto an interacting Schr\"{o}dinger
equation. It might be possible to solve this equation exactly too,
using techniques developed recently for the Sutherland
Hamiltonian.\cite{Sim93} From the work of Chalker and
Mac\^{e}do\cite{Cha93} we know that the two-point correlation function
of the eigenvalues in the large-$N$ limit has a simple
$1/\beta$-dependence on the symmetry index. This poses strong
restrictions on a possible $\beta$-dependence of the eigenvalue
interaction, which can only differ from the form (\ref{final3u})
derived for $\beta=2$ on intervals $\Delta\lambda\simeq L/Nl\ll 1$
comparable to the spacing between the $\lambda$'s.

It might be possible to come up with another maximum-entropy principle,
different from that of Muttalib, Pichard, and Stone,\cite{Mut87} which
yields the correct eigenvalue interaction (\ref{final3u}) instead of
the logarithmic interaction (\ref{Pglobalb}). Slevin and
Nagao\cite{Sle93} have proposed an alternative maximum-entropy
principle, but their distribution function does not improve the
agreement with Eq.\ (\ref{UCF}). It would be particularly worthwhile to
find an intuitive explanation for the halving of the logarithmic
interaction for weakly transmitted scattering channels.

To go beyond quasi-one-dimensional geometries (long and narrow wires)
remains an outstanding problem. A numerical study of Slevin, Pichard,
and Muttalib\cite{Sle93b} has indicated a significant break-down of the
logarithmic repulsion for two- and three-dimensional geometries
(squares and cubes). A generalization of the DMPK equation (\ref{DMPK})
to higher dimensions has been the subject of some recent
investigations.\cite{Mel91b,Cha93b} It remains to be seen whether the
method reviewed here for Eq.\ (\ref{DMPK}) is of use for that problem.

\nonumsection{Acknowledgements}
\noindent
The research reviewed in this paper was carried out in collaboration
with B. Rejaei. It was supported financially by the ``Ne\-der\-land\-se
or\-ga\-ni\-sa\-tie voor We\-ten\-schap\-pe\-lijk On\-der\-zoek'' (NWO)
and by the ``Stich\-ting voor Fun\-da\-men\-teel On\-der\-zoek der
Ma\-te\-rie'' (FOM).

\nonumsection{References}

\end{document}